\documentclass[lettersize,journal]{IEEEtran}

\usepackage{amsmath,amssymb,amsfonts}
\usepackage{algorithmic}
\usepackage{algorithm}
\usepackage{array}
\usepackage[caption=false,font=normalsize,labelfont=sf,textfont=sf]{subfig}
\usepackage{textcomp}
\usepackage{stfloats}
\usepackage{url}
\usepackage{verbatim}
\usepackage{graphicx}
\usepackage{cite}
\usepackage[capitalize,nameinlink,compress]{cleveref}
\usepackage{enumitem}
\usepackage{mathtools}
\usepackage{pgfplots}
\pgfplotsset{compat=newest}
\usetikzlibrary{plotmarks}
\usetikzlibrary{arrows.meta}
\usepgfplotslibrary{patchplots}
\usepackage{grffile}
\usepackage{amsmath}

\newcommand{\R}{\mathbb{R}}

\makeatletter
\newcommand{\pushright}[1]{\ifmeasuring@#1\else\omit$\displaystyle#1$\ignorespaces\fi}
\makeatother

\usepackage[prependcaption,colorinlistoftodos]{todonotes}

\newtheorem{problem}{Problem}

\crefname{appendix}{Appendix}{Appendices}
\crefname{figure}{Figure}{Figures}
\crefname{line}{line}{lines}
\crefname{claim}{Claim}{Claims}
\crefname{equation}{}{}
\crefname{problem}{Problem}{Problems}
\crefname{assumption}{Assumption}{Assumptions}

\newcommand{\markAsChanged}[1]{{\color{black}#1}}

\begin{document}

\title{Safety Filter for Limiting the Current of Grid-Forming Matrix Modular Multilevel Converters}

\author{Michael Schneeberger, Silvia Mastellone, Florian D{\"o}rfler 
}

\maketitle

\begin{abstract}
Grid-forming (GFM) converters face significant challenges in limiting current during transient grid events while preserving their grid-forming behavior.
This paper offers an elegant solution to the problem with a priori guarantees, presenting a safety filter approach based on Control Barrier Functions (CBFs) to enforce current constraints with minimal deviation from the nominal voltage reference.
The safety filter is implemented as a Quadratic Program, enabling real-time computation of safe voltage adjustments that ensure smooth transitions and maintain the GFM behavior during nominal operation. 
To provide formal safety certificate, the CBF is synthesized offline using a Sum-of-Squares optimization framework, ensuring that the converter remains within its allowable operating limits under all conditions.
Additionally, a Control Lyapunov Function is incorporated to facilitate a smooth return to the nominal operating region following grid events. 
The proposed method is modular and can be integrated into many of the GFM control architectures, as demonstrated with two different GFM implementations. 
High-fidelity simulations conducted with an enhanced matrix modular multilevel converter connected to both high-inertia and low-inertia grid scenarios validate the effectiveness of the safety filter, showing that it successfully limits current during faults, preserves GFM behavior, and ensures a seamless recovery to nominal operation.
\end{abstract}

\section{Introduction}

In low-inertia power grids, such as microgrids, railways and industrial grids, grid-forming (GFM) converters are essential for maintaining a stable grid voltage, allowing grid-following (GFL) converters to synchronize and sustain stable internal DC-link voltages.
However, compared to synchronous machines, GFM converters can support only smaller current amplitudes and rely on protective mechanisms to prevent hardware damage during overcurrent events.
If these protection mechanisms are triggered, the converter may abruptly shut down, leading to operational unavailability, significant financial losses, and damaged equipment.
Consequently, the converter current must be carefully controlled to prevent such scenarios.
This requirement, however, introduces a seeming contradiction: GFM converters must simultaneously exhibit inertial behavior to ensure grid stability under normal operating conditions, while also responding rapidly to grid transients -- similar to the behavior of GFL converters -- to limit current and prevent converter shutdown.
Addressing this trade-off is critical to ensuring the reliable and efficient operation of GFM converters in both high-inertia and low-inertia power systems.

Various Current Limiting Control (CLC) strategies have been developed to restrict converter current during grid transients.
One common approach involves directly regulating the current via a Current Controller (CC).
For instance, in \cite{rocabert2012control}, a nested voltage and current controller is employed to regulate the voltage across the capacitor, assuming the presence of a sufficiently large capacitor at the converter's output.
In other methods \cite{huang2021impact}, the current reference of the current controller is determined from a voltage source behind impedance model, with the voltage being set by the GFM control.
While a permanently activated current controller performs well in high-inertia grids, it can lead to instabilities due to racing conditions with other current control based converters \cite{taul2019current}.
One potential solution to this issue is to deactivate the current control loop under nominal operating conditions and activate it only when the current approaches its threshold \cite{ndreko2018gfm_paper}.
However, this method introduces undesirable transients caused by the on-off switching of the current controller.
A smoother alternative is to use a proportional current controller with current-dependent activation, as proposed in \cite{wu2024design}. 
While this method avoids abrupt transitions, the lack of integral action can result in steady-state errors \cite{baeckeland2024overcurrent}, potentially undermining the GFM converter's ability to maintain grid stability during fault conditions.
Another method involves directly computing a dynamic voltage drop using an adaptive virtual impedance model \cite{wu2021small}, effectively reducing the terminal voltage during faults to limit the current.
However, this model only captures the steady-state relation between voltage and current, but fails to account for the dynamical behavior of the system.

While these methods offer practical solutions for current limitation in some scenarios, they often require cumbersome manual tuning tailored to each specific system architecture and range of worst-case operating scenarios.
If tuned improperly, they can undermine the GFM behavior of the converter even during nominal conditions.
Furthermore, a significant limitation of these techniques is the lack of formal safety certification. 
None of these methods guarantee that the system will remain within its allowable operating limits under all conditions.

In this paper, we propose a safety filter approach based on a Control Barrier Function (CBF) to limit the converter current.
CBFs are a widely recognized tool for ensuring the forward invariance of dynamical systems, with manifold applications in the robotics and learning communities.
The safe converter voltage is computed by solving online a Quadratic Program (QP), which minimizes deviations from the GFM voltage reference.
The QP constraints encode the CBF condition, ensuring that voltage adjustments are only applied as the system approaches high-current conditions, while preserving the intended GFM behavior during normal operation.
The safety filter is a modular component that can be integrated to any GFM control architecture, which we exemplify with two different GFM control schemes; Virtual Synchronous Machine (VSM) and Enhanced Direct Power Control (EDPC).
\markAsChanged{A key advantage of this method, beyond its favorable performance, is the ability to provide an a priori formal certificate of safety, eliminating the need for extensive simulations and control tuning to verify safety.
This certificate is established during the design phase using a Sum-of-Squares (SOS) approach when searching for a polynomial CBF.}
Additionally, the safety filter can be augmented with a Control Lyapunov Function (CLF), as demonstrated in our previous work \cite{schneeberger2024advanced}, to guarantee a smooth return to the nominal operating region after a grid event.
This improves the transition between current limiting operation and nominal GFM operation, and guarantees inactivity of the safety filter within a nominal region, thereby automating the tuning of the safety filter.
The resulting QP can be solved explicitly and does not require a numerical solver \cite{ames2016control}, making the safety filter easy to implement in real-world applications.
To validate the effectiveness of the safety filter, we conduct simulations in both high-inertia and low-inertia grid scenarios, involving an advanced Matrix Modular Multilevel Converter (M3C) model.
The performance of our safety filter is compared to alternative current limiting controllers.
These simulations highlight the applicability, efficacy, and favorable performance of our safety filter concept across different GFM control and realistic grid scenarios.

The remainder of the paper is structured as follows:
\cref{sec:gfm_control} provides an overview of two GFM control schemes and examines various current limiting strategies.
In \cref{sec:safty_filter}, we present our proposed safety filter approach for limiting the current of a GFM converter.
\cref{sec:test_cases} details test setup used to evaluate the proposed method.
\cref{sec:test_results} demonstrates the effectiveness of the safety filter through numerical results and comparisons with alternative approaches.
Finally, \cref{sec:conclusion} summarizes the key finding and contributions of the paper.

\section{Challenges of GFM Converters} \label{sec:gfm_control}

\begin{figure}
   \centering
   \resizebox{86mm}{!}{\includegraphics{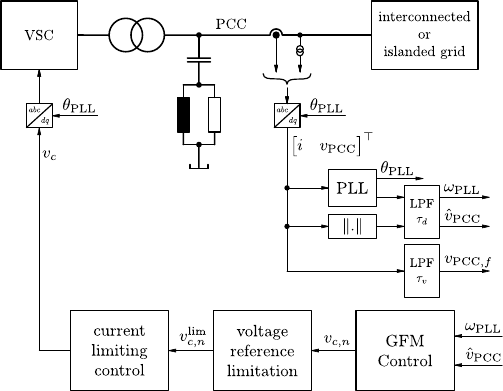}}
   \caption{
      The GFM converter setup consists of a Voltage Source Converter (VSC) connected to the PCC through a transformer.
      The grid filter suppresses high-frequency harmonics generated by the modulated converter voltage.
   }
   \label{fig:converter_schematic}
\end{figure}

In this section, we present two GFM control schemes -- Virtual Synchronous Machine (VSM) and Enhanced Direct Power Control (EDPC) --, and examine various current-limiting strategies.
\cref{fig:converter_schematic} illustrates the setup, which also contains a voltage reference limitation, which ensures a steady-state converter current within the constraints.
Throughout the paper, any current $i := \begin{bmatrix}
   i_d & i_q
\end{bmatrix}^\top$ or voltage $v := \begin{bmatrix}
   v_d & v_q
\end{bmatrix}^\top$ is expressed in the dq reference frame.
The zero component of the current is denoted by $i_0$.
All parameters are given in per-unit unless specified otherwise.
Both GFM control schemes adapt the converter power $p$ and converter frequency $\omega$ in steady state according to a frequency droop characteristic.
This is done by first defining a power reference $p_r$ (that later enters the main GFM control loop):
\begin{align} \label{def:inverse_frequency_droop}
   (p_r - p^*) = (-1/D_f) (\omega_\text{PLL} - \omega^*),
\end{align}
where $D_f$ is the frequency droop constant,
$\omega_\text{PLL}$ is the filtered PLL frequency,
$p^*$ is the active power set point, and
$\omega^*$ is the frequency set point.
The set points $p^*$ and $\omega^*$ allow for adjustments to variable steady-state and dispatch conditions; for our purposes, they are kept constant at $p^* = 0$ p.u. and $\omega^* = 1$ p.u.

\subsection{Virtual Synchronous Machine (VSM)}

\begin{figure}
   \centering
   \subfloat[\label{fig:vsm_control_1}]{\resizebox{85mm}{!}{\includegraphics{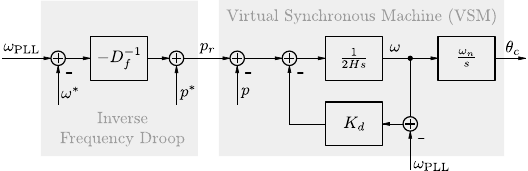}}}
   \vfil
   \subfloat[\label{fig:edpc_1}]{\resizebox{85mm}{!}{\includegraphics{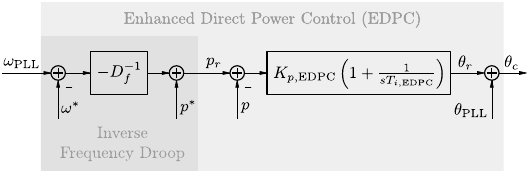}}}
   \caption{
      Two GFM control strategies: (a) the VSM, as defined in \cref{def:vsm}, emulates the physical behavior of a SM by dynamically integrating the converter's output power, and (b) the EDPC, as defined in \cref{def:edpc}, combines the phase of the PLL with the output of an active power loop. For both the power reference is determined by the inverse frequency droop in \cref{def:inverse_frequency_droop}.
   }
   \label{fig:vsm_control}
\end{figure}

The VSM, as illustrated in \cref{fig:vsm_control}(a), emulates the physical behavior of a Synchronous Machine (SM) by dynamically integrating the converter's output power according to \cite{d2013virtual}:
\begin{align} \label{def:vsm}
   2H \frac{d \omega_c}{dt} = (p_r - p) - K_d (\omega_c - \omega_\text{PLL}),
\end{align}
where $\omega_c$ is the converter frequency,
$H$ is the inertia constant (in seconds), and
$K_d$ is the damping constant.
The power reference $p_r$ from \cref{def:inverse_frequency_droop} serves as an analog to the mechanical power in a SM determined by a turbine-governor.
The second term in the swing equation \cref{def:vsm} acts as a damping mechanism, mitigating frequency oscillations that arise when the converter is interconnected with other components governed by the same swing equation dynamics.
Under steady-state conditions, where $\frac{d\omega_c}{dt} = 0$ and $\omega_c = \omega_\text{PLL}$, the power $p$ matches the power reference $p_r$, thereby satisfying the frequency droop relationship $(p - p^*) = (-1/D_f) (\omega_\text{PLL} - \omega_c^*)$.

\subsection{Enhanced Direct Power Control (EDPC)}

The EDPC, as depicted in \cref{fig:vsm_control}(b), leverages the inertia-like behavior of the PLL combined with an active power loop, as defined in \cite{rosso2021grid}:
\begin{align} \label{def:edpc}
   \theta_c = \theta_\text{PLL} + \underbrace{K_{p,\text{EDPC}}  \left ( 1 + \frac{1}{s T_{i, \text{EDPC}}} \right ) (p_r - p)}_{=: \theta_r},
\end{align}
where 
$\theta_\text{PLL}$ is the phase of the PLL,
$\theta_c$ is the phase of the converter's terminal voltage,
$K_{p,\text{EDPC}}$ and $T_{i, \text{EDPC}}$ are proportional and integral gains of the PI controller, and
$p_r$ is the reference power derived from the inverse frequency droop in \cref{def:inverse_frequency_droop}.
Unlike for the VSM approach, the inverse frequency droop is an integral part of the EDPC, as shown in \cref{fig:vsm_control}(b).
The output of the active power loop, denoted by $\theta_r$, defines the difference between the phase of the PLL and the phase of the terminal voltage.

As highlighted in \cite{schweizer2022grid}, a properly tuned PLL can replicate certain characteristics of the swing equation.
By coupling the terminal voltage phase directly to the PLL, the EDPC dynamically adjust the converter frequency in a manner analogous to a SM.
To illustrate this mechanism, consider the scenario where a constant current flows through the equivalent transformer inductance $Z_c$ into the GFM converter.
In this situation, the phase difference between the terminal voltage and the voltage at PCC, denoted as $\theta_{Z_c} = \theta_c - \theta_\text{PLL}$, remains constant.
Consequently, the PLL control error simplifies to $\theta_\text{PCC} - \theta_\text{PLL} = \theta_r - \theta_{Z_c}$.
This demonstrates that $\theta_r$ (apart from a constant offset $\theta_{Z_c}$) directly modulates the PLL frequency, thereby mimicking the inertial response of a SM.

\subsection{Voltage amplitude control and voltage reference limitation} \label{sec:voltage_limitation}

\begin{figure}
   \centering
   \resizebox{86mm}{!}{\includegraphics{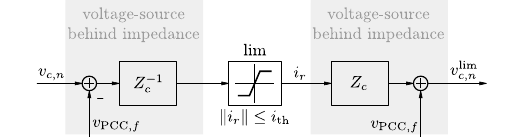}}
   \caption{
      illustrates the implementation of voltage reference limitation using a voltage behind an impedance model.
      This approach ensures that a current reference amplitude remains below the current threshold, i.e. $\|i_r\| \leq i_\text{th}$.
   }
   \label{fig:voltage_reference_limitation}
\end{figure}

The amplitude of the converter voltage is determined using a reactive power controller based on the voltage droop characteristics $(\hat v_\text{PCC} - v^*) = D_v (q_r - q^*)$, where the set points are fixed at $q^* = 0$ p.u. and $v^* = 1$ p.u.
Furthermore, to ensure a stable steady-state current, the GFM voltage reference $v_{c,n}$ is limited, as illustrated in \cref{fig:converter_schematic}.
This is achieved by generating a current reference $i_r$ based on a voltage source behind an impedance model, as illustrated in \cref{fig:voltage_reference_limitation}, using the equivalent transformer impedance:
\begin{align*}
   Z_c := r_c + J \omega l_c,
\end{align*}
where $J = \begin{bsmallmatrix}0 & -1 \\ 1 & 0\end{bsmallmatrix}$, and $\omega$ is the p.u. angular frequency.
The precise value of the resistance $r_c$ and inductance $l_c$ are typically determined during the commissioning of the GFM converter.
The current reference is constrained by the inequality $\| i_r \|^2 \leq i_{\text{th}}^2$, where $i_{\mathrm{th}}$ represents a predefined current threshold.
This limitation prioritizes the direct current component $i_d$ over the quadrature current component $i_q$, emulating the cross-forming behavior presented in \cite{he2024cross}.
The limited converter voltage reference is then computed as:
\begin{align} \label{def:limited_voltage_reference}
   v_{c,n}^\text{lim} = Z_c i_r + v_\text{PCC,f}.
\end{align}
Here, the low-pass filtered PCC voltage $v_{\text{PCC},f}$ (with time constant $\tau_v$) is used instead of the direct PCC voltage measurement $v_{\text{PCC}}$ to reduce the converter's sensitivity to rapid grid voltage fluctuations.

\subsection{Conventional current-limiting control strategies for GFM controls} \label{sec:current_limiting_strategies}

\begin{figure}
   \centering
   \subfloat[\label{fig:current_control}]{\resizebox{86mm}{!}{\includegraphics{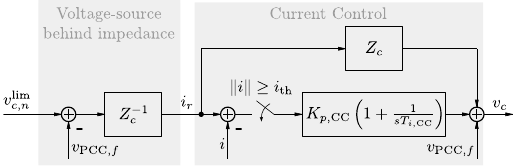}}}
   \vfil
   \subfloat[\label{fig:clc}]{\resizebox{86mm}{!}{\includegraphics{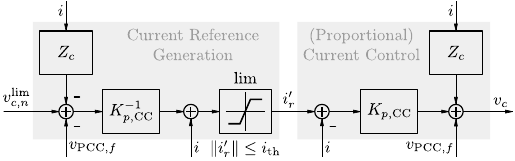}}}
   \vfil
   \subfloat[\label{fig:adaptive_vi}]{\resizebox{86mm}{!}{\includegraphics{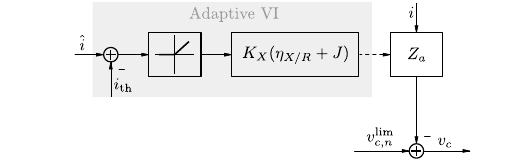}}}
   \caption{Current-limiting control strategies: (a) Switched Current Control (SCC), (b) Reference-Limited Proportional Current Control (RL-CC), and (c) Adaptive Virtual Impedance (AVI).}
   \label{fig:current_limiting_control}
\end{figure}

The GFM converter current must be limited during transient grid event to protect the system's hardware and prevent operational loss.
Limiting the current of a GFM converter presents two seemingly conflicting requirements for the CLC:
On the one hand, it must closely follow the GFM voltage reference $v_{c,n}$ to provide a stable reference for GFL converters.
On the other hand, it must rapidly adjust the converter voltage -- akin to a GFL converter -- to limit the current, regardless of the GFM voltage reference.
To reconcile these seemingly conflicting requirements, the CLC, as illustrated in \cref{fig:converter_schematic}, must be designed to intervene only when necessary, ensuring that the current limitation is achieved with minimal disruption of the converter's GFM operation.
In the following, we present three state-of-the-art current-limiting control approaches that address this challenge:

\begin{enumerate}[label=\alph*)]
   \item A \textit{Switched Current Control} (SCC) \cite{ndreko2018gfm_paper}, as depicted in \cref{fig:current_limiting_control}(a) relies on a conventional current controller with proportional gain $K_{p,\text{CC}}$ and time constant $T_{i,\text{CC}}$.
   The PI controller of the SCC is activated -- temporarily shifting the converter into GFL operation -- whenever the current exceeds its threshold, i.e. $\| i \| \geq i_\text{th}$, ensuring that current limitation is applied only when required.
   To mitigate extensive switching, a hysteresis mechanism can be introduced.
   The SCC is an effective method to switch between GFM and GFL operation.
   However, due to the oscillatory nature of the current, SCC often requires frequent activation, leading to undesirable transients in the system.
   \item A \textit{Reference-Limited Proportional Current Control} (RL-CC) \cite{wu2024design}, as illustrated in \cref{fig:current_limiting_control}(b), provides an alternative to the switched current controller by employing a proportional current controller with gain $K_{p,\text{CC}}$.
   Unlike SCC, RL-CC enables smooth activation based on whether a fictitious current reference $i_r'$ exceeds the threshold $i_\text{th}$, thereby reducing the transients caused by abrupt switching.
   However, the absence of integral action can lead to a steady-state current error in high-current conditions, causing a persistent deviation from the GFM voltage reference.
   \item Another approach to reducing the current amplitude is by using an \textit{Adaptive Virtual Impedance} (AVI) \cite{wu2021small}, as depicted in \cref{fig:current_limiting_control}(c).
   This technique introduces a dynamic voltage drop by applying a virtual impedance, which becomes active when the current exceeds the threshold $i_\text{th}$. 
   The inductive component of the virtual impedance is calculated as the product of the positive deviation of the current amplitude from $i_\text{th}$ and a constant $K_X$, while the resistive component is derived from the virtual impedance ratio $\eta_{X/R}$.
   The AVI method effectively limits the current by reducing the converter voltage through this virtual impedance. However, since the voltage source behind the impedance model only captures the steady-state voltage-current relationship, it does not fully account for the system's dynamic behavior.
\end{enumerate}

While these methods offer viable solutions for current limitation in certain cases, they require cumbersome manual tuning for specific, often tailored to worst-case scenarios, lack verification by extensive simulations, and lack formal safety certificates. 
This risks current limit violations during unforeseen conditions, potentially triggering protection mechanisms, underscoring the need for more robust and certifiable approaches.

\section{Safety Filter based on CBF} \label{sec:safty_filter}

\begin{figure}
   \centering
   \resizebox{86mm}{!}{\includegraphics{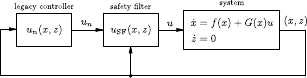}}
   \caption{
      The safety filter is a modular component that can be integrated into the existing control hierarchy, ideally without compromising the performance of the legacy controller during nominal operation.
      State with zero time derivative, referred to as stationary states, are grouped into a separate variable $z$.
   }
   \label{fig:safety_filter_schematic}
\end{figure}

To address the limitations of the state-of-the-art CLC, we propose a safety filter approach to smoothly limit the current of a GFM converter.
The modular nature of the safety filter enables to elegantly integrate it into the existing control hierarchy (see \cref{fig:safety_filter_schematic}), ideally without compromising the performance of the legacy controller during nominal operation.
It operates based on the concept of a \textit{safe set} $\mathcal X_s$ (see \cref{fig:safe_and_nominal_set}), which is required to be a subset of the allowable set $\mathcal X_a$ encoding the current limits.
By ensuring forward invariance of this set, the safety filter guarantees that the converter remains within its current limits at all times, without unnecessary interference during nominal conditions.
Additionally, an enhanced variant incorporating a CLF, as detailed in our previous work \cite{schneeberger2024advanced}, enables a smooth return to \textit{nominal region} $\mathcal X_n$ following a grid disturbance.

\subsection{Specifications for safety filter design}

\begin{figure}
   \centering
   \resizebox{68mm}{!}{\includegraphics{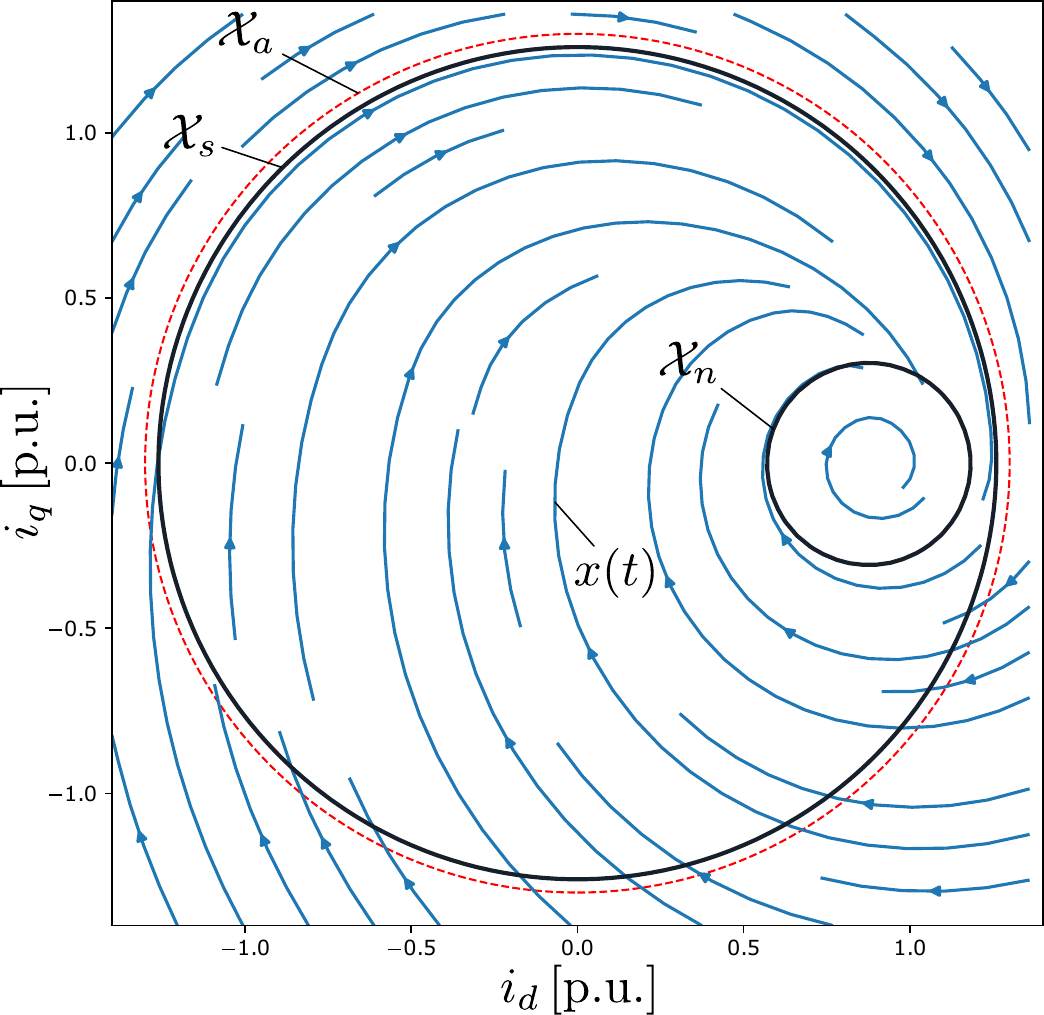}}
   \caption{
      illustrates the vector field $\dot x = f(x) + G u_\text{SF}(x,z)$ in blue, representing the converter dynamics in \cref{eq:simplified_system}, where the stationary state $z$ is composed of
      $i_r = \begin{bmatrix}0.9 & 0\end{bmatrix}^\top$, $\Delta v_{\text{PCC},f} = \begin{bmatrix}0 & 0\end{bmatrix}^\top$, and $i_0 = 0$.
      The safe set $\mathcal X_s = \{ x \mid B(x, z) \leq 0 \}$ specifies a forward invariant set that is contained in the allowable set $\mathcal X_a = \left \{ x \mid i_{d}^2 + i_{q}^2 \leq i_\mathrm{max}^2 \right \}$, encoding the maximum allowed current constraints.
      The CLF condition ensures a finite-time convergence to the nominal region $\mathcal X_n = \{ x \mid V(x, z) \leq 0 \}$.
   }
   \label{fig:safe_and_nominal_set}
\end{figure}

The safety filter is designed to adapt to the dynamical nature of the grid, which can exhibit arbitrarily complex behavior and evolve over time.
Moreover, the nominal GFM control scheme in \cref{sec:gfm_control} may exhibit an unrequired degree of complexity for the safety filter design.
To ensure computational tractability, we introduce simplifications based on the assumption that certain signals change slowly relative to the time-scale of the safety filter, allowing us to approximate them as constant:
\begin{enumerate}
   \item \textbf{PCC voltage $v_\text{PCC}$}: 
   Assuming a stationary PCC voltage allows us to express the internal dynamical model of the safety filter in terms of the filtered PCC voltage deviations $\Delta v_{\text{PCC},f} = v_{\text{PCC},f} - v_{\text{PCC}}$, reducing the number of state variables.
   \item \textbf{Zero-component current $i_0$}:
   Together with the current $i$, the zero-component current $i_0$ determines the worst-case current amplitude in each of the three phases, captured in the allowable set $\mathcal X_a$.
   \item \textbf{Current reference $i_r$}: 
   The nominal region $\mathcal X_n$ in the $(i_d, i_q)$ plane is defined based on this reference (see \cref{fig:safe_and_nominal_set}).
\end{enumerate}
When applying the safety filter later, real-time measured values are used, even though they are treated as constant within the internal dynamical model of the safety filter.

Given the assumptions of a stationary $v_\text{PCC}$, 
the current dynamics across the transformer and output filter, along with the dynamics of the filtered PCC voltage deviation are defined as:
\begin{subequations} \label{eq:simplified_system}
   \begin{align}
      \dot x &= \underbrace{\begin{bmatrix}
         -(\omega_n / l_c) \left( Z_c i \right) \\
         -(1/\tau_v) \Delta v_{\mathrm{PCC},f}
      \end{bmatrix}}_{=:f(x)} + \underbrace{\begin{bmatrix}
         (\omega_n / l_c) \mathrm{I}_2 \\ \mathrm{0}
      \end{bmatrix}}_{=:G(x)} u, \\
      \dot z &= 0
   \end{align}
\end{subequations}
where the state is partitioned into non-stationary state $x := \begin{bmatrix}i^\top & \Delta v_{\text{PCC},f}^\top \end{bmatrix}^\top$ and stationary state $z := \begin{bmatrix}i_r^\top & i_0 \end{bmatrix}^\top$, as depicted in \cref{fig:safety_filter_schematic}.
The system input is defined in terms of the voltage difference $u := v_c - v_\text{PCC}$, where $v_c$ is the converter's terminal voltage regulated by the safety filter.

\markAsChanged{Within the nominal region $\mathcal X_n$, the safety filter's output should align with the GFM voltage reference defined in \cref{def:limited_voltage_reference}, referred to as the nominal control.
Since $i_r$ is assumed stationary, the nominal control depends only on $(x, z)$, eliminating the need to model additional GFM control dynamics.}
Based on this, we define the nominal controller as:
\begin{align} \label{eq:nominal_control_action}
   u_n(x, z) := i_r Z_c + \Delta v_{\text{PCC},f}.
\end{align}
Furthermore, by assuming $i_0$ is constant, the allowable set $\mathcal X_a$ can be explicitly formulated in terms of $(x, z)$.
Since current limitation must be enforced in all three phases, not just the dq reference frame, the allowable set is given as:
\begin{align} \label{eq:allowable_set}
   \mathcal X_a := \left \{ (x,z) \mid \underbrace{-i^\top i + (i_\text{max} - i_0)^2}_{=: w(x,z)} \leq 0 \right \}.
\end{align}
The detailed derivation is provided in the appendix in \cref{sec:worst_case_abc}.

Finally, the system input is constrained by the maximum voltage the converter can apply at its terminal, which depends on the converter topology and the total DC-link voltage.
To maintain generality, the maximum converter voltage amplitude is approximated by a constant maximum modulation index $m_\text{max} \in \R$ (assuming a nominal DC-link voltage).
Under the assumption of a nominal PCC voltage $v_{\text{PCC},n} = \begin{bmatrix}1 & 0\end{bmatrix}^\top$, the input constraint set is expressed as:
\begin{align} \label{eq:input_constraint_set}
   \mathcal U := \{ u \mid (u + v_{\text{PCC},n})^\top (u + v_{\text{PCC},n}) \leq m_\text{max} \},
\end{align}
where $m_\text{max}$ represents the maximum modulation index independent of the topology.

\subsection{Control Barrier and Lyapunov-like Function and problem formulation}

The safe set is defined as the zero-sublevel set $\mathcal X_s := \{ x \mid B(x,z) \leq 0 \}$, which must be contained within the allowable set $\mathcal X_a$.
The forward invariance of the safe set $\mathcal X_s$ is established through the scalar polynomial $B(x, z)$.
A polynomial satisfying the regularity condition in \cite[Definition 4.9]{Blanchini:99} is termed a \underline{Control Barrier Function} (CBF) if, for all states at the boundary of the safe set $x \in \partial \mathcal X_s$, there exists an input $u \in \mathcal U$ such that (cf. \cite[Nagumo's Theorem 4.7]{Blanchini:99}):
\begin{align} \label{eq:cbf_condition}
    \nabla_x B(x,z)^\top \left ( f(x) + G(x) u \right ) \leq 0.
\end{align}

To facilitate a smooth return to the nominal region after a grid disturbance, we introduce a scalar polynomial $V(x, z)$, referred to as \underline{Control Lyapunov-like Function} (CLF) that measures the distance to the nominal region.
Such a finite convergence is achieved if, for all states in the transitional region \markAsChanged{$x \in \mathcal X_t := \{ x \mid B(x, z) \leq 0 \leq V(x,z) \}$}, there exists and input $u \in \mathcal U$ such that $V(x(t))$ is strictly decreasing (cf. \cite{schneeberger2024advanced}):
\begin{align} \label{eq:clf_condition}
    \dot V(x, z) = \nabla_x V(x, z)^\top \left ( f(x) + G(x) u \right ) \leq - d(x, z).
\end{align}
The nominal region is defined as the zero-sublevel set $\mathcal X_n := \{ x \mid V(x, z) \leq 0 \}$, which must be contained within the safe set $\mathcal X_s$.
A strictly positive dissipation rate $d(x, z)$ within $\mathcal X_t$ guarantees finite-time convergence to $\mathcal X_n$.
To ensure compatibility between the CBF and CLF, conditions \cref{eq:cbf_condition,eq:clf_condition} must hold for the same input $u \in \mathcal U$.
Additionally, we impose a constraint on the CLF: for all states at the boundary of the nominal region $x \in \partial \mathcal X_n$, the following must hold:
\begin{align} \label{eq:clf_condition_un}
    \nabla_x V(x, z)^\top \left ( f(x) + G(x) u_n(x, z) \right ) + d(x, z) \leq 0.
\end{align}
\markAsChanged{This ensures that the system remains within the nominal region $\mathcal X_n$ when using the nominal control $u_n(x, z)$ in \cref{eq:nominal_control_action}, allowing the safety filter to remain inactive during nominal operation.}

The problem around finding $B(x, z)$ and $V(x, z)$ that fulfill the CBF and CLF condition can be formulated as follows:
\begin{problem} \label{prob:cbf_and_clf}
   Given the dynamical system \cref{eq:simplified_system}, the legacy controller $u_n(x, z)$ in \cref{eq:nominal_control_action}, the allowable set of states $\mathcal X_a$ in \cref{eq:allowable_set}, and the input constraint set $\mathcal U$ in \cref{eq:input_constraint_set}, find functions $B(x, z)$ and $V(x, z)$ such that: \begin{itemize}
      \item (forward-invariance of $\mathcal X_s$) CBF and CLF conditions \cref{eq:cbf_condition,eq:clf_condition} are satisfied for the same input $u \in \mathcal U$,
      \item (forward-invariance of $\mathcal X_n$ w.r.t. $u_n(x, z)$) the nominal region condition \cref{eq:clf_condition_un} is satisfied,
      \item (set containment) the containment conditions $\mathcal X_n \subsetneq \mathcal X_s \subsetneq \mathcal X_a$ are satisfied,
      \item (objective) CBF $B(x, z)$ maximizes the volume of the safe set, denoted by $\text{vol}(\mathcal X_s)$.
   \end{itemize}
\end{problem}
\markAsChanged{The search for compatible CBF $B(x, z)$ and CLF $V(x, z)$ in the space of polynomials using SOS optimization tools is presented in the appendix \cref{sec:sos_optimization}.}

\subsection{Quadratic Program (QP)}

Once feasible CBF and CLF candidates are identified, the safety filter can be implemented using a Quadratic Program (QP).
Given the real-time measurement of the non-stationary state $x$ and stationary state $z$, the safety filter control law $u_\text{SF}(x,z)$ is computed at each time step of the discrete controller by solving a QP.
\markAsChanged{The QP finds the least modification of the nominal control $u_n(x, z)$ to ensure both the forward invariance of the safe set $\mathcal X_s$, as enforced by the CBF $B(x, z)$, and the finite-time convergence to the nominal region $\mathcal X_n$, as dictated by the CLF $V(x, z)$:}
\begin{align} \label{eq:qp_based_controller}
   \begin{array}{lll}
      \underset{u \in \R^2}{\mbox{min}} & \| u_n(x, z) - u \|^2 \\
      \mbox{s.t.} & \nabla_x B(x,z)^\top \left ( G(x) u + f(x) \right ) \leq -\gamma_B B(x,z) \\
      &\nabla_x V(x,z)^\top \left ( G(x) u + f(x) \right ) \leq -\gamma_V V(x,z),
   \end{array}
\end{align}
where $G(x)$ and $f(x)$ are taken from the system in \cref{eq:simplified_system}.
The values of $\gamma_B > 0$ and $\gamma_V > 0$ are set equal to the (constant) SOS polynomials that emerge in the quadratic module of \cref{eq:sos_constr_ineq_cbf,eq:sos_constr_ineq_clf} with respect to $B(x,z)$ and $V(x,z)$.
To simplify the implementation on a controller, we omit the input constraints in the online optimization problem.
\cref{fig:safe_and_nominal_set} provides a visual representation of the safe set $\mathcal X_s$, and the nominal set $\mathcal X_n$, along with the vector field of system \cref{eq:simplified_system} operating the safety filter \cref{eq:qp_based_controller}.
The complete implementation, including the optimization and simulation code, is available online\footnote{https://github.com/MichaelSchneeberger/advanced-safety-filter-mmc/}.

\section{Description of Test Cases} \label{sec:test_cases}

\begin{figure}
   \centering
   \resizebox{80mm}{!}{\includegraphics{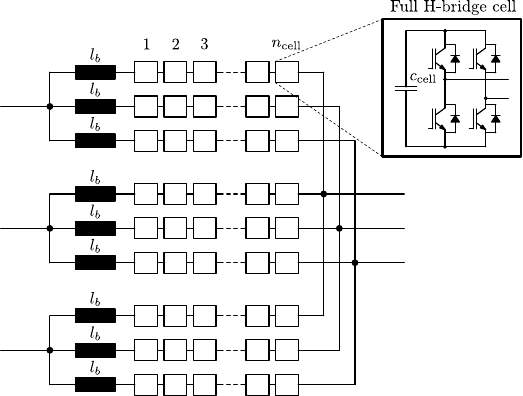}}
   \caption{
      Matrix Modular Multilevel Converter (M3C) topology consists of nine branches, each configured with a branch inductance $l_b$ and $n_\text{cell}$ cells.
      Each cell within the branches comprises a full H-bridge and a capacitor $c_\text{cell}$.
   }
   \label{fig:mmc_schematic}
\end{figure}

\begin{figure}
   \centering
   \subfloat[\label{fig:sim_setup_2}]{\resizebox{85mm}{!}{\includegraphics{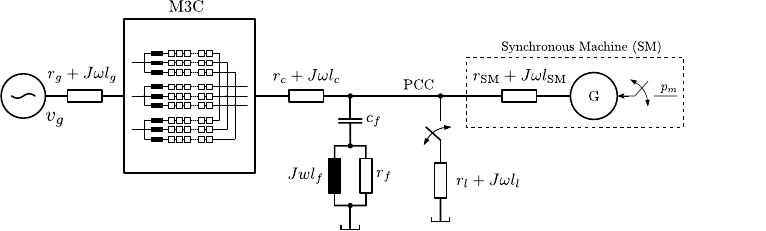}}}
   \vfil
   \subfloat[\label{fig:sim_setup_1}]{\resizebox{85mm}{!}{\includegraphics{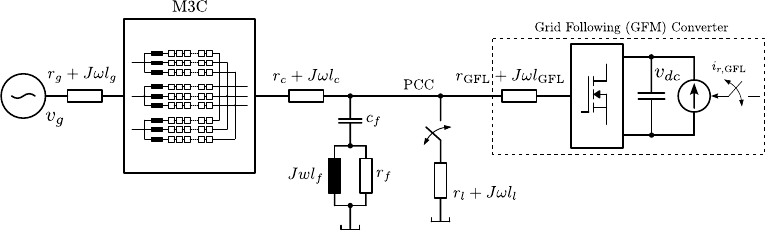}}}
   \caption{A worst-case grid fault at the M3C converter's terminal is performed by the short circuit switch in (a) a high-inertia grid implemented through SM, and (b) a low-inertia grid using a GFL converter.}
   \label{fig:sim_setup}
\end{figure}

The performance of the safety filter is evaluated against the current-limiting strategies presented in \cref{sec:current_limiting_strategies}.
For this purpose, we utilize an advanced Matrix Modular Multilevel Converter (M3C) model, as depicted in \cref{fig:mmc_schematic}, that connects the utility grid, represented by a stiff grid, to an industrial grid.
The M3C topology consists of nine branches, each configured with a branch inductance $l_b$ and $n_\text{cell}$ cells.
Each cell within the branches comprises a capacitor $c_\text{cell}$ and a full H-bridge, with the simulation model explicitly representing the dyanamics of all four switches.
A GFL control scheme is employed on the utility grid side to ensure a stable total DC-link voltage (given by the sum of all cell voltages), while a GFM control scheme is employed on the industrial grid side to stabilize the voltage.
Worst-case grid faults are simulated by switching a small inductance directly at the PCC, as depicted in \cref{fig:sim_setup}.
These faults are applied for 300ms in both a high-inertia and low-inertia industrial grid.
The performance of the CLC are assessed based on three key metrics: the maximum current overshoot, the magnitude of the converter voltage deviation $\Delta v := v_c - v_{c,n}^\text{lim}$ applied by the CLC, as well as current stability.
All simulation parameters are summarized in \cref{tab:param_values}.

The nine branch currents of the M3C are analyzed using the two-state $\alpha \beta 0$ transformation, as detailed in \cite{kammerer2012fully}. 
Two out of the nine transformed currents, corresponding to the $\alpha \beta$ currents of the utility grid, are regulated using a standard current control strategy.
Four of the nine transformed currents, referred to as the circulating currents, are controlled based on a current controller, as detailed in \cite{diaz2017control}.
One of the nine transformed current corresponds to the zero current component, which does not require active control.
The remaining two transformed currents represent the $\alpha \beta$ currents of the industrial grid and are regulated using either of the two GFM controls, presented in \cref{sec:gfm_control}, combined with a CLC method.
The various CLC presented in \cref{sec:current_limiting_strategies} are tuned to best limit the current while not disturbing the nominal operation.
All the controls are executed discretely with a sampling time of 200 $\mu s$.

We consider two grid configurations, one representing a high-inertia grid and one representing a low-inertia grid.
\begin{enumerate}
   \item \textbf{High-inertia grid}: This grid is implemented using a single synchronous machine (SM) connected to the M3C converter, as depicted in \cref{fig:sim_setup}(b).
   The SM is modelled by a voltage source with voltage amplitude fixed at 1p.u. and the voltage phase given by the swing equation:
   \begin{align}
      2H \frac{d \omega_\text{SM}}{dt} = (p_m - p_\text{SM}),
   \end{align}
   where $\omega_\text{SM}$ is the SM's mechanical frequency, $p_m$ is the mechanical power, and $p_\text{SM}$ is the electrical output power.
   When the short-circuit occurs, the mechanical power is reset to zero to enable resynchronization of the GFM converter with the SM after the fault is cleared.
   \item \textbf{Low-inertia grid}: For this scenario, a single GFL converter, representing an aggregation of smaller GFL converter, is connected to the GFM converter, as depicted in \cref{fig:sim_setup}(a).
   The GFL converter employs an outer DC-link voltage control loop, providing a current reference to the inner current controller.
   An internal PLL provides the phase for the dq reference frame of the GFL converter control.
   A current source attached to the DC-link of the GFL converter injects a current $i_{r,\text{GFL}}$, representing the active power consumed by a load or produced by a source.
   In such a configuration, the current flowing from the GFM converter into the grid is determined in steady-state by the current reference $i_{r,\text{GFL}}$ of the GFL converter.
   When the short-circuit occurs, the current reference $i_{r,\text{GFL}}$ is reset to zero to maintain nominal DC-link voltage.
\end{enumerate}
The low-inertia and high-inertia scenarios described above represent two limit cases, each highlighting distinct control challenges.
In the low-inertia scenario, the output power is primarily determined by the GFL converter, while the GFM control is responsible for regulating the grid frequency.
Conversely, in the high-inertia scenario, the grid frequency is dictated by the synchronous machines, leaving the GFM control to manage the output power.
Testing the control strategies in these distinct scenarios provides valuable insights into their performance across a wide range of grid configurations, where both the output power and the grid frequency vary dynamically.

\begin{table}[!t]
   \caption{Simulation parameters}
   \label{tab:param_values}
   \centering
   \begin{tabular}{|c|c|c|c|} 
      \hline
      Parameter & Symbol & Value & Unit \\
      \hline \hline
      Utility grid & $l_g$ & 0.32 & p.u. \\
      & $r_g$ & 0.02 & p.u. \\
      Transformer & $l_c$ & 0.16 & p.u. \\
      & $r_c$ & 0.02 & p.u. \\
      Grid filter & $c_f$ & 0.006 & p.u. \\
      & $l_f$ & 0.2 & p.u. \\
      & $r_f$ & 10 & p.u. \\
      SM impedance & $l_\text{SM}$ & 0.16 & p.u. \\
      & $r_\text{SM}$ & 0.01 & p.u. \\
      GFL VSC impedance & $l_\text{GFL}$ & 0.16 & p.u. \\
      & $r_\text{GFL}$ & 0.01 & p.u. \\
      Load & $l_l$ & 0.016 & p.u. \\
      & $r_l$ & 0.001 & p.u. \\
      \# Cells per branch & $n_\text{cell}$ & 14 & - \\
      Cell capacitor & $c_\text{cell}$ & 5 & mF \\
      Cell voltage & $v_\text{cell}$ & 5 & kV \\
      Branch inductance & $l_b$ & 5& mH \\
      & $r_b$ & 10 & m$\Omega$ \\
      Grid frequency & $f$ & 60 & Hz \\
      Maximum current & $i_\mathrm{max}$ & 1.30 & p.u. \\
      Current threshold & $i_{\mathrm{th}}$ & 1.18 & p.u. \\
      Frequency droop & $D_f$ & 0.02 & p.u. \\
      Voltage droop & $D_v$ & 0.05 & p.u. \\
      AVI gain & $K_X$ & 10 & p.u. \\
      X/R ratio & $\eta_{X/R}$ & 16 & - \\
      Max. voltage deviation & $\Delta v_{\text{PCC},f,\text{max}}$ & 1 & p.u. \\
      Max. current ref. amplitude & $i_{r,\text{max}}$ & 1.18 & p.u. \\
      Max. zero-comp. current & $i_{0,\text{max}}$ & 0.6 & p.u. \\
      Dissipation rate & $d_r$ & 0.1 & - \\
      QP constants & $\gamma_B$ & 211 & - \\
      & $\gamma_V$ & 683 & - \\
      CC PI control & $K_{p,\text{CC}}$ & 0.342 & p.u. \\
      & $T_{i, \text{CC}}$ & 2 & ms \\
      PLL PI control & $K_{p,\text{PLL}}$ & 0.096 & p.u. \\
      & $T_{i, \text{PLL}}$ & 85 & ms \\
      EDPC PI control & $K_{p,\text{EDPC}}$ & 0.45 & p.u. \\
       & $T_{i, \text{EDPC}}$ & 120 & ms \\
      LPF time constant & $\tau_d$ & 10 & ms \\
      LPF time constant & $\tau_v$ & 100 & ms \\
      Damping Constant & $K_d$ & 50 & - \\
      Inertia Constant & $H$ & 3 & s \\
      Mechanical Power & $p_m$ & 0.9 & p.u. \\
      Current Reference & $i_{r, \text{GFL}}$ & -0.9 & p.u. \\
      \hline
   \end{tabular}
\end{table}

\section{Results and Analysis}  \label{sec:test_results}

\markAsChanged{This section presents a comparison of conventional CLCs, including the Switched Current Control (SCC), Reference-Limited Proportional Current Control (RL-CC), and the Adaptive Virtual Impedance (AVI), against our proposed safety filter approach.}

\subsection{Case Study \romannumeral1: CLC in high-inertia grid}

\begin{figure*}[!t]
   \centering
   \subfloat[\label{fig:sm_vsm1} SCC]{\resizebox{44mm}{!}{\includegraphics{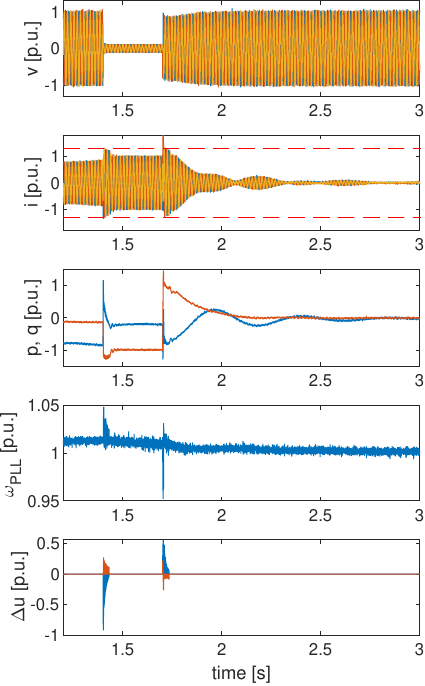}}}
   \hfil
   \subfloat[\label{fig:sm_vsm2} RL-CC]{\resizebox{44mm}{!}{\includegraphics{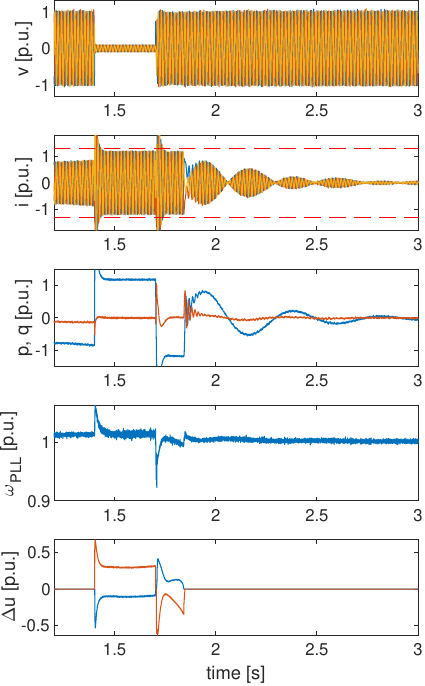}}}
   \hfil
   \subfloat[\label{fig:sm_vsm3} AVI]{\resizebox{44mm}{!}{\includegraphics{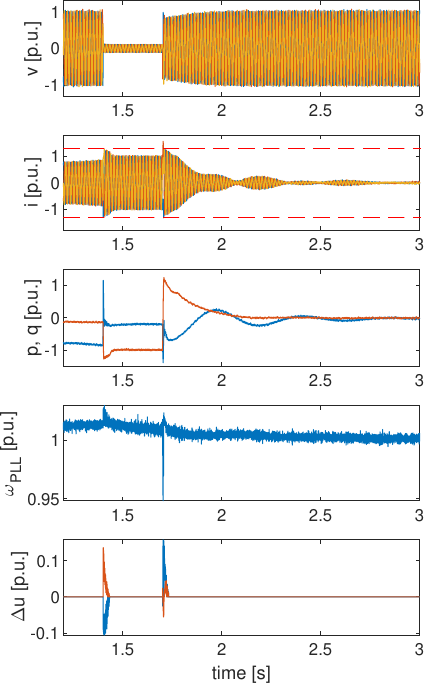}}}
   \hfil
   \subfloat[\label{fig:sm_vsm4} Safety Filter]{\resizebox{44mm}{!}{\includegraphics{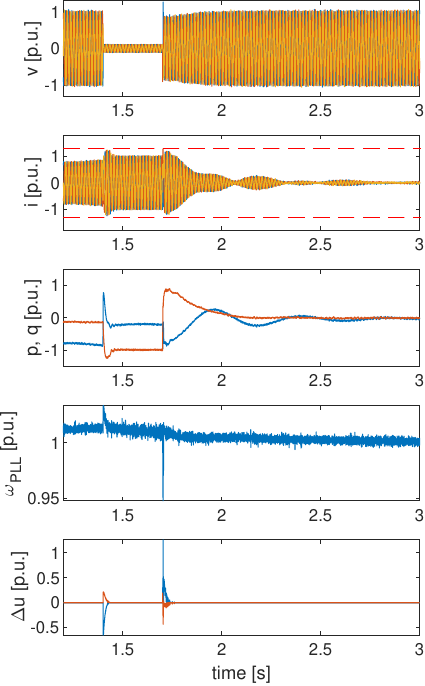}}}
   \caption{High inertia case study: A grid fault is introduced at the PCC of the GFM converter, which operates in conjunction with a SM using the VSM control strategy and equipped with the current-limiting control.}
   \label{fig:sm_vsm}
\end{figure*}

\begin{figure*}[!t]
   \centering
   \subfloat[\label{fig:sm_edpc1} SCC]{\resizebox{44mm}{!}{\includegraphics{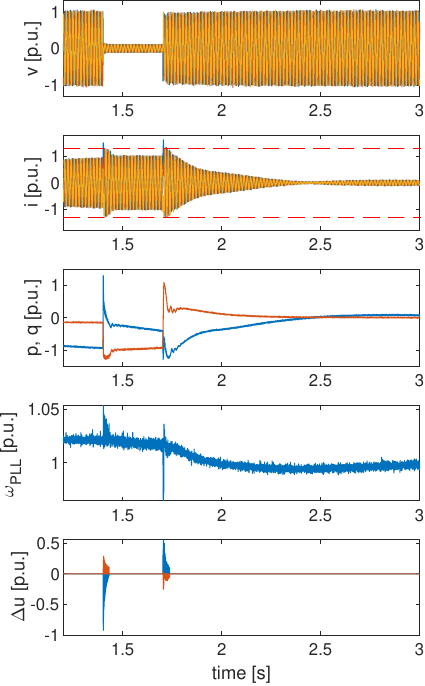}}}
   \hfil
   \subfloat[\label{fig:sm_edpc2} RL-CC]{\resizebox{44mm}{!}{\includegraphics{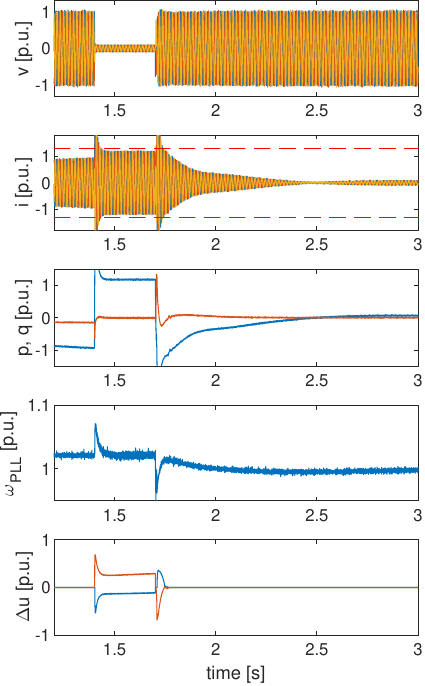}}}
   \hfil
   \subfloat[\label{fig:sm_edpc3} AVI]{\resizebox{44mm}{!}{\includegraphics{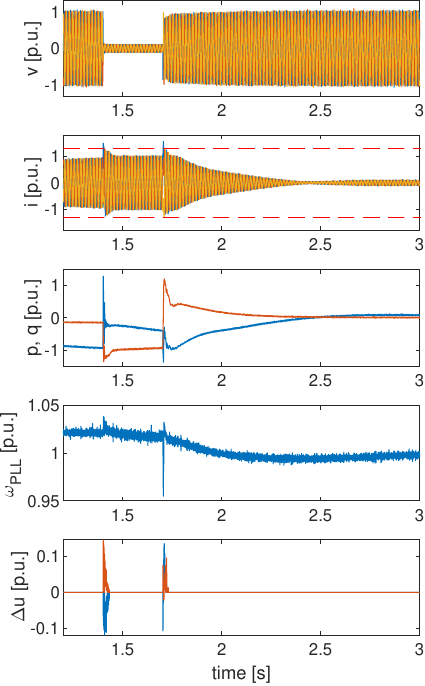}}}
   \hfil
   \subfloat[\label{fig:sm_edpc4} Safety Filter]{\resizebox{44mm}{!}{\includegraphics{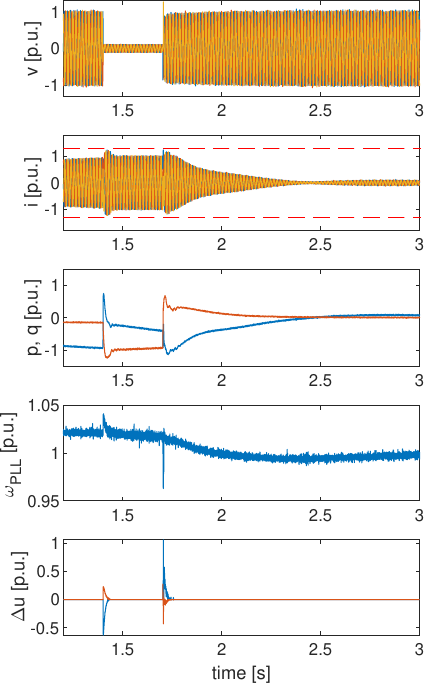}}}
   \caption{High inertia case study: A grid fault is introduced at the PCC of the GFM converter, which operates in conjunction with a SM using the EDPC strategy and equipped with the current-limiting control.}
   \label{fig:sm_edpc}
\end{figure*}

In this case study, the conventional CLC strategies, described in \cref{sec:current_limiting_strategies}, and our proposed safety filter are evaluated in the high-inertia grid configuration introduced in the previous section.
The results are depicted in \cref{fig:sm_vsm} for the VSM and in \cref{fig:sm_edpc} for the EDPC.
Apart from the RL-CC, it is observed that all CLC strategies intervene only during the fault event, with no converter voltage deviation $\Delta v$ applied during nominal operation.
The RL-CC strategy offsets the voltage reference during the fault, failing to properly track the GFM control action.
When the fault clears, maximum negative active power is injected through the GFM converter for another 100ms, until the RL-CC ceases to intervene.
Furthermore, the conventional SCC, RL-CC, and AIV methods exceed the current limit both when the fault occurs and when it is cleared.
While none of the tested conventional CLC strategies ensures safety, our proposed safety filter effectively limits the current with both GFM control approaches.
Moreover, the intervention $\Delta u$ introduced by the safety filter remains smooth and minimal, ensuring precise and reliable current limitation without disrupting the GFM behavior.
Beyond its favorable performance, the safety filter stands out as the only CLC strategy with a priori theoretical guarantees, i.e., we specify and guarantee its favorable performance already in the design phase.

\begin{figure}[!t]
   \centering
   \subfloat[\label{fig:sm_vsm_comp1}]{\resizebox{44mm}{!}{\includegraphics{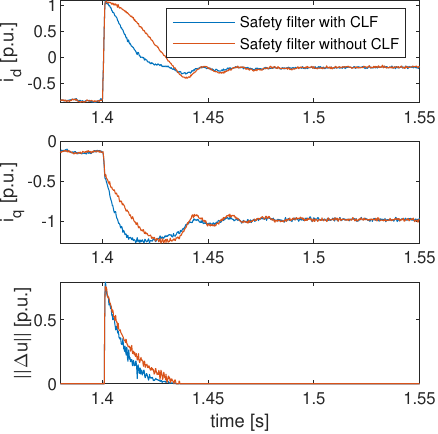}}}
   \hfil
   \subfloat[\label{fig:sm_vsm_comp2}]{\resizebox{44mm}{!}{\includegraphics{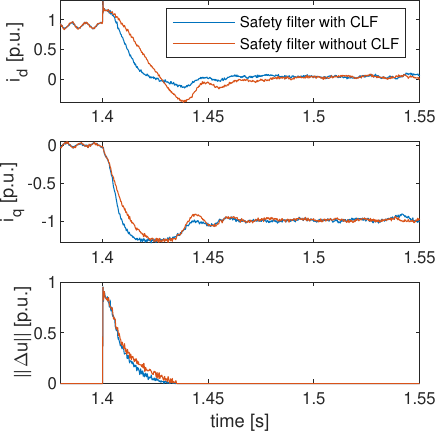}}}
   \caption{
      illustrates the performance of the safety filter in two configurations: with CLF condition and without CLF condition, applied to (a) a high-inertia grid, and (b) a low-inertia grid.
      The safety filter including the CLF condition exhibits a faster recovery to the nominal operation immediately after the fault occurs.
   }
   \label{fig:sm_vsm_comp}
\end{figure}

\cref{fig:sm_vsm_comp1} compares the dynamic response of a safety filter with and without a CLF condition when the grid fault is applied in a high-inertia grid.
The results show that incorporating a CLF accelerates transient recovery and slightly reduces current oscillations, involving less intervention $\Delta u$.
This leads to a faster retracking of the GFM voltage reference after the fault.

\subsection{Case Study \romannumeral2: CLC in low-inertia grid}

\begin{figure*}[!t]
   \centering
   \subfloat[\label{fig:gfl_vsm1} SCC]{\resizebox{44mm}{!}{\includegraphics{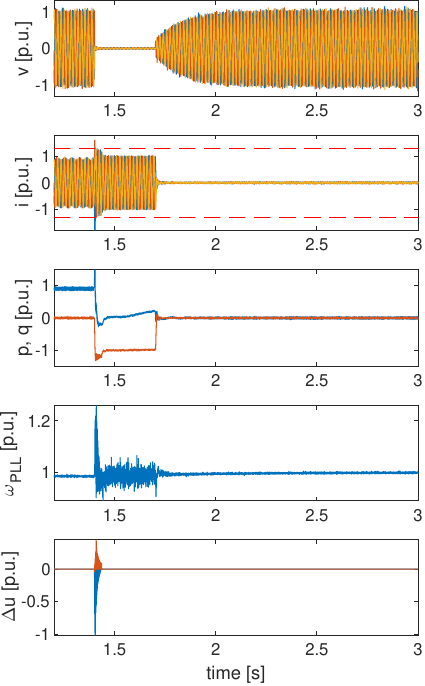}}}
   \hfil
   \subfloat[\label{fig:gfl_vsm2} RL-CC]{\resizebox{44mm}{!}{\includegraphics{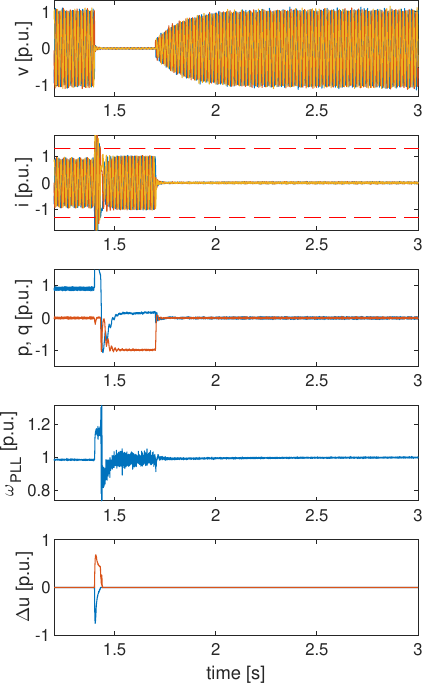}}}
   \hfil
   \subfloat[\label{fig:gfl_vsm3} AVI]{\resizebox{44mm}{!}{\includegraphics{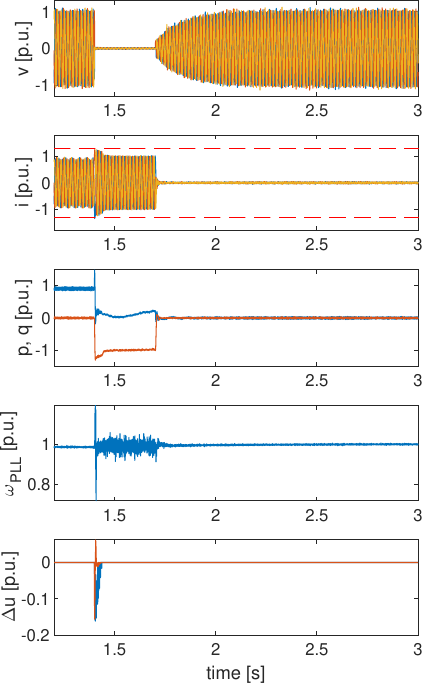}}}
   \hfil
   \subfloat[\label{fig:gfl_vsm4} Safety Filter]{\resizebox{44mm}{!}{\includegraphics{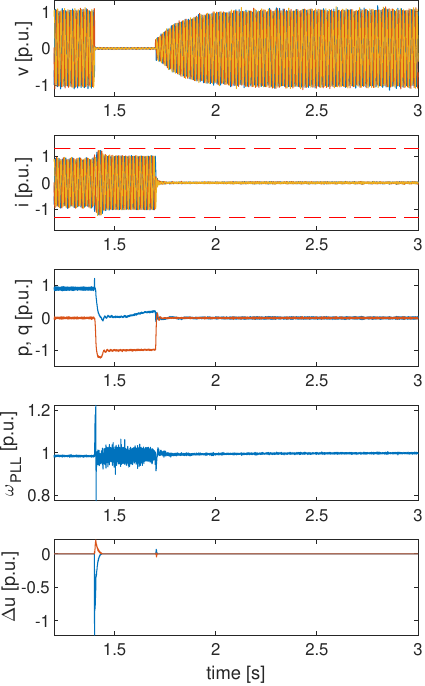}}}
   \caption{Low inertia case study: A grid fault is introduced at the PCC of the GFM converter, which operates in conjunction with a GFL converter using the VSM control strategy and equipped with the current-limiting control.}
   \label{fig:gfl_vsm}
\end{figure*}

\begin{figure*}[!t]
   \centering
   \subfloat[\label{fig:gfl_edpc1} SCC]{\resizebox{44mm}{!}{\includegraphics{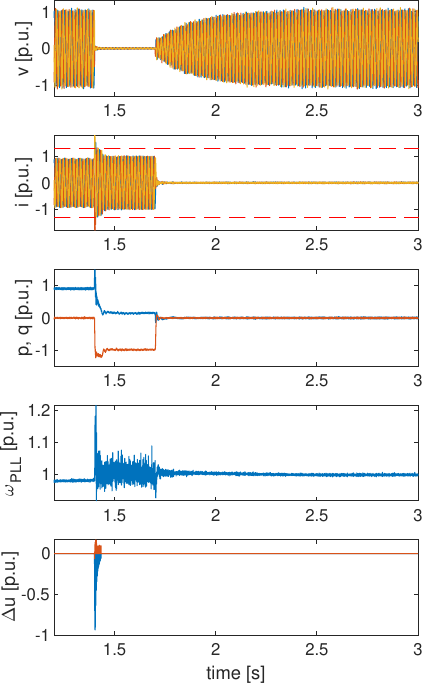}}}
   \hfil
   \subfloat[\label{fig:gfl_edpc2} RL-CC]{\resizebox{44mm}{!}{\includegraphics{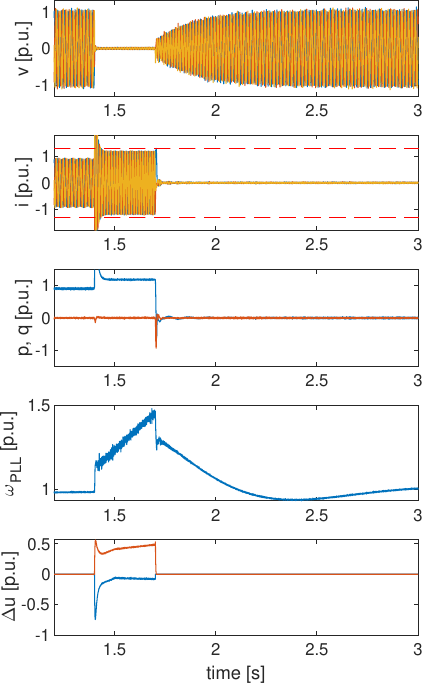}}}
   \hfil
   \subfloat[\label{fig:gfl_edpc3} AVI]{\resizebox{44mm}{!}{\includegraphics{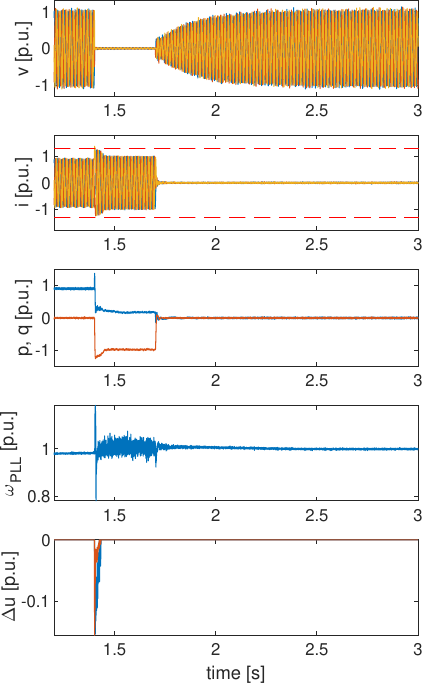}}}
   \hfil
   \subfloat[\label{fig:gfl_edpc4} Safety Filter]{\resizebox{44mm}{!}{\includegraphics{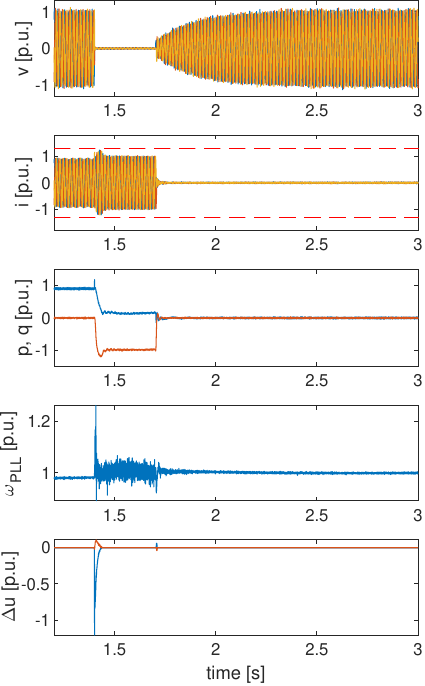}}}
   \caption{Low inertia case study: A grid fault is introduced at the PCC of the GFM converter, which operates in conjunction with a GFL converter using the EDPC strategy and equipped with the current-limiting control.}
   \label{fig:gfl_edpc}
\end{figure*}

In this case study, the conventional CLC strategies and our safety filter are evaluated in the low-inertia grid configuration introduced in the previous section.
The results are depicted in \cref{fig:gfl_vsm} for the VSM and in \cref{fig:gfl_edpc} for the EDPC.
All CLC strategies intervene exclusively during the fault event, with no converter voltage deviation $\Delta u$ applied during nominal operation.
Again, the RL-CC strategy offsets the voltage reference during the fault, which prevents proper tracking of the GFM control action.
When employing the EDPC, the converter injects maximum active power into the grid during the fault, leading to a continuous increase in the grid frequency.
Furthermore, the SCC, RL-CC, and AIV methods fail to maintain the current within the maximum allowed current when the fault occurs.
None of the tested conventional CLC strategies from \cref{sec:current_limiting_strategies} ensure safety under all conditions. 
In contrast, our proposed safety filter reliably limits the current with both GFM control methods. Furthermore, the intervention $\Delta v$ applied by the safety filter is both smooth and minimal, providing effective current limitation while preserving the GFM behavior without disruptions.
Again, all of these favorable properties are guaranteed a priori in the design of our safety filter.

\cref{fig:sm_vsm_comp2} compares the dynamic response of a safety filter with and without a CLF condition during a grid fault in a low-inertia grid.
The results demonstrate that integrating a CLF enhances transient recovery and marginally reduces current oscillations, necessitating less intervention $\Delta u$.
This leads to a faster retracking of the GFM voltage reference after the fault.

\section{Conclusions} \label{sec:conclusion}

In this paper, we presented and demonstrated a novel safety filter approach for current limitation in GFM converters.
The proposed method was rigorously tested in a realistic setup involving the complex topology of a Matrix Modular Multilevel Converter (M3C) connected to both high-inertia and low-inertia grids. 
Our results show that the advanced safety filter effectively limits current during grid faults while preserving the GFM behavior of the converter, as exemplified by its compatibility with both the Virtual Synchronous Machine (VSM) and the Enhanced Droop Control (EDPC) schemes.
The performance of our proposed solution is superior compared to other CLC methods and guaranteed a priori tuning the design of the safety filter.
This approach ensures reliable and safe operation under transient conditions, highlighting its potential for enhancing the stability and resilience of modern power systems.

\section{Appendix}

\subsection{Worst-case current in abc frame} \label{sec:worst_case_abc}

We claim that the condition $x \in \mathcal X_a$ \cref{eq:allowable_set} ensures that the current in each phase remains within the maximum current value.
Specifically, we need to show that for any current vector $i_{dq0} = \begin{bmatrix}i_d & i_q & i_0\end{bmatrix}^\top = \begin{bmatrix} \hat i cos(\phi) & \hat i sin(\phi) & i_0 \end{bmatrix}^\top$ that satisfies the condition defining $\mathcal X_a$:
\begin{align*}
   i^\top i = (\hat i)^2 \leq (i_\text{max} - i_0)^2,
\end{align*}
then also $i_{abc} := \begin{bmatrix}i_a & i_b & i_c\end{bmatrix}^\top \leq i_\text{max}$.
First, we note that the inverse Park transformation is given as:
\begin{align*}
   K_p^{-1}(\theta) = \begin{bmatrix}
      cos(\theta) & -sin(\theta) & 1 \\
      cos(\theta-\frac{2\pi}{3}) & -sin(\theta-\frac{2\pi}{3}) & 1 \\
      cos(\theta+\frac{2\pi}{3}) & -sin(\theta+\frac{2\pi}{3}) & 1. \\
   \end{bmatrix}
\end{align*}
The angle $\phi$ that results in the worst-case current amplitudes of $\pm \hat i + i_0$ in phase a is given by $\phi = \mp \theta$.
The same worst-case current amplitudes are achieved for the other two phases for an angle shifted by $\pm 2 \pi /3$.
Therefore, the current amplitudes can be bounded as:
\begin{align*}
   i_{abc} = K_p^{-1}(\theta) i_{dq0} \leq \hat i + i_0 \leq i_\text{max}.
\end{align*}
The second inequality follows from the definition of $\mathcal X_a$.

\subsection{SOS Optimization} \label{sec:sos_optimization}

The coefficients of the polynomial CBF $B(x, z)$ and CLF $V(x, z)$, as specified in \cref{prob:cbf_and_clf}, are determined numerically by formulating the inequality conditions \cref{eq:cbf_condition,eq:clf_condition,eq:clf_condition_un}, the input constraints $u \in \mathcal U$, and the containment conditions $\mathcal X_n \subsetneq \mathcal X_s \subsetneq \mathcal X_a$ as \textit{Sums-of-Squares} (SOS) constraints using \textit{Putinar's Positivstellensatz}.
SOS is a powerful mathematical optimization technique used to certify the non-negativity of polynomials by expressing them as a sum of squared polynomials $p(x) = \sum_k \left ( q_k(x) \right )^2 \in \Sigma[x]$.
This approach is widely used in control theory to formulate non-negativity constraints over $\R^n$, which can then be translated into constraints of a convex optimization problem.

Furthermore, the non-negativity constraint can also be established on a subset $\mathcal X \subseteq \R^n$ defined by the intersection of zero-sublevel sets of polynomials $f_1, ..., f_{n_c} \in R[x]$ through the SOS constraint:
\begin{align*}
   p(x) - \gamma_1(x) f_1(x) - ... - \gamma_{n_c}(x) f_{n_c}(x) \in \Sigma[x],
\end{align*}
where $\gamma_1, ..., \gamma_{n_c} \in \Sigma[x]$ are auxiliary polynomials whose coefficients are determined via the SOS optimization problem.
Putinar's Positivstellensatz guarantees that a strictly positive solution on $\mathcal X$ also satisfy this SOS constraint, though without specifying the degree of the auxiliary polynomials.

To ensure compatibility between the CBF, CLF, and the input constraints, the control input $u$ in \cref{eq:cbf_condition,eq:clf_condition} is replaced with a polynomial control law $u_\mathrm{SOS}(x,z)$.
Additionally, an auxiliary polynomial variable $v$ is introduced to explicitly encode the input constraints, as detailed in \cite{schneeberger2024advanced}.
The resulting SOS constraints are defined over the polynomial variables $(x, z, v)$:
\begin{subequations} \label{eq:sos_constr_ineq}
   \begin{align}
      \begin{split} \label{eq:sos_constr_ineq_cbf}
         -\nabla_x B^\top \left ( f + G u_\mathrm{SOS} \right ) &\in \textbf{M}(B, -B, -f_{\mathrm{op}, 1 .. 4}) 
      \end{split} \\
      \begin{split} \label{eq:sos_constr_ineq_clf}
         -\nabla_x V^\top \left ( f + G u_\mathrm{SOS} \right ) - d &\in \textbf{M}(V, -B, -f_{\mathrm{op}, 1 .. 4}) 
      \end{split} \\
      \begin{split} \label{eq:sos_constr_ineq_nom}
         -\nabla_x V^\top \left ( f + G u_{n}' \right ) - d &\in \textbf{M}(V, -V, -f_{\mathrm{op}, 1 .. 4}),
      \end{split} \\
      \begin{split}
         -v^\top (u_\mathrm{SOS} + v_{\text{PCC},n}) + m_\text{max} & \in \textbf{M}(-B, -f_u, -f_{\mathrm{op}, 1 .. 4})
      \end{split} \\
      \begin{split}
         B & \in \textbf{M}(w, -f_{\mathrm{op}, 2 .. 4})
      \end{split}\\
      \begin{split}
         V & \in \textbf{M}(B, -f_{\mathrm{op}, 2 .. 4}),
      \end{split}
   \end{align}
\end{subequations}
where a \emph{quadratic module} generated by polynomials $f_1, ..., f_{n_c} \in R[x]$ is defined by
\begin{align}
   \textbf M(f_1, ..., f_{n_c}) := \left \{ \gamma_0 + \sum_{i=1}^{n_c} \gamma_i f_i \mid \gamma_0, ..., \gamma_{n_c} \in \Sigma[x] \right \}.
\end{align}
The dissipation rate is given by $d(x, z) = d_r (V(x, z) + \epsilon)$ for some $d_r > 0$ and a small $\epsilon > 0$, and the function encoding the input constraints in $v$ is given as $f_u(v) = v^\top v - m_\text{max}$.
For simplicity, function arguments are omitted in the notation in \cref{eq:sos_constr_ineq}.
To ensure that the SOS constraints account for realistic operating condition, we define an operational region $\mathcal X_\mathrm{op}$ that captures the permissible ranges of both stationary and non-stationary states:
\begin{align}
   \mathcal X_\text{op} := \left \{ (x, z) \mid f_{\mathrm{op}, k}(x,z) \leq 0 \text{ for } k = 1, ..., 4 \right \},
\end{align}
where
\begin{align*}
   f_{\mathrm{op}, 1}(x,z) &= w(x, z) \\
   f_{\mathrm{op}, 2}(x,z) &= -\Delta v_{\text{PCC},f}^\top \Delta v_{\text{PCC},f} + \Delta v_{\text{PCC},f,\text{max}}^2 \\
   f_{\mathrm{op}, 3}(x,z) &= -i_r^\top i_r + (i_{r,\text{max}} - i_0)^2 \\
   f_{\mathrm{op}, 4}(x,z) &= -(i_0-i_{0,\text{max}}/2)^2 + (i_{0,\text{max}}/2)^2.
\end{align*}
Building upon our previous work \cite{schneeberger2024advancedbess}, the nominal controller in \cref{eq:sos_constr_ineq_nom} was refined during the offline design phase.
The modified controller is given by:
\begin{align*}
   u_{n}'(x, z) := 0.2 (i_r - i) + u_n(x, z).
\end{align*}
Using a refined nominal controller compromises the forward invariance of the nominal region.
However, without this adjustment, the optimization would either be infeasible or yield a nominal region $\mathcal X_n$ that is too small, leading to an overly restrictive safety filter. 
A small nominal region would cause unnecessary interventions even during nominal operation, contradicting the design goal of a minimally intrusive safety filter.

The final SOS optimization problem is formulated as:
\begin{align} \label{eq:sos_opt_prob}
   \hspace{-\leftmargin}
   \begin{array}{ll}
      \mbox{find} & V, B, u_\mathrm{SOS} \\
      \mbox{maximize} & \text{vol}(\mathcal X_s) + \lambda \text{vol}(\mathcal X_n) \\
      \mbox{subject to} & \text{\cref{eq:sos_constr_ineq}} \text{ and } B(0, 0) = -1.
   \end{array}
\end{align}
where $vol(\mathcal X)$ is the volume of the set $\mathcal X$.
Maximizing the volume of the safe set $\mathcal X_s$ and the nominal region $\mathcal X_n$ ensures safety for wide range of system states while minimizing the intervention of the safety filter, thereby preserving the nominal control action for a large set of operating conditions.
The hyperparameter $\lambda \geq 0$ allows for a tunable trade-off between the volume of $\mathcal X_s$ and $\mathcal X_n$.
The constant part of $B(x,z)$ is fixed to $-1$ to ensure numerical stability when solving the SOS problem.
The solution is obtained through an iterative approach, alternating between searching over one set of decision variables while keeping the others fixed.
For a detailed discussion of the solution methodology, we refer to the supplementary material in \cite{schneeberger2024advanced}.
The computed coefficients of the polynomials CBF and CLF are presented in \cref{eq:cbf_parameters} and \cref{eq:clf_parameters}.

\begin{figure*}[ht]
   \normalsize
   \begin{align} \label{eq:cbf_parameters}
      B(x, z) =  0.63 i_d^2 + 0.63 i_q^2 - 0.63 i_0^2 + 1.59 i_0 - 1
   \end{align}
   \begin{align} \label{eq:clf_parameters}
      \begin{split}
      V(x, z) = 
      &4.70 i_d^2 
      - 13.98 i_d \Delta v_{\text{PCC},f,d} 
      - 30.29 i_d \Delta v_{\text{PCC},f,q} 
      - 9.15 i_d i_{r,d}
      - 0.02 i_q i_{r,d}
      + 4.70 i_q^2 
      + 30.29 i_q \Delta v_{\text{PCC},f,d} \\
      &+ 13.98 i_q \Delta v_{\text{PCC},f,q}
      + 0.02 i_q i_{r,d}
      - 9.15 i_q i_{r,q}
      + 115.48 \Delta v_{\text{PCC},f,d}^2
      + 15.73 \Delta v_{\text{PCC},f,d} i_{r,d}
      - 33.67 \Delta v_{\text{PCC},f,d} i_{r,q} \\
      &+ 115.48 \Delta v_{\text{PCC},f,q}^2
      + 33.67 \Delta v_{\text{PCC},f,q} i_{r,d}
      + 15.73 \Delta v_{\text{PCC},f,q} i_{r,q}
      + 5.15 i_{r,d}^2
      + 5.15 i_{r,q}^2
      - 0.63 i_0^2 
      + 1.59 i_0 
      - 1
      \end{split}
   \end{align}
   \hrulefill 
   \caption{The coefficients of the polynomials CBF $B(x, z)$ and CLF $V(x, z)$ are computed using the SOS optimization problem in \cref{eq:sos_opt_prob}.}
\end{figure*}

\bibliographystyle{unsrt}
\bibliography{main}

\vskip -2\baselineskip plus -1fil

\end{document}